\documentclass{article}

\usepackage{arxiv}
\usepackage[utf8]{inputenc} 
\usepackage[T1]{fontenc}    
\usepackage{hyperref}       
\usepackage{url}            
\usepackage{booktabs}       
\usepackage{amsfonts}       
\usepackage{nicefrac}       
\usepackage{microtype}      
\usepackage{lipsum}
\usepackage{graphicx}
\graphicspath{ {./images/} }

\title{Evaluating Voice Command Pipelines for Drone Control: From STT and LLM to Direct Classification and Siamese Networks}

\author{
 Lucca Emmanuel Pineli Simões \\
  Instituto de Informática (INF)\\
  Universidade Federal de Goiás\\
  Goiânia, Brazil \\
  \texttt{lucca.pineli@discente.ufg.br} \\
  \And
 Lucas Brandão Rodrigues \\
  Instituto de Informática (INF)\\
  Universidade Federal de Goiás\\
  Goiânia, Brazil \\
  \texttt{brandao.brandao@discente.ufg.br} \\
  \And
 Rafaela Mota Silva \\
  Instituto de Informática (INF)\\
  Universidade Federal de Goiás\\
  Goiânia, Brazil \\
  \texttt{rafaelamota@discente.ufg.br} \\
  \And
 Gustavo Rodrigues da Silva \\
  Instituto de Informática (INF)\\
  Universidade Federal de Goiás\\
  Goiânia, Brazil \\
  \texttt{rodrigues\_da@discente.ufg.br} \\
}

\begin{document}
\maketitle
\begin{abstract}
This paper presents the development and comparative evaluation of three voice command pipelines for controlling a Tello drone, using speech recognition and deep learning techniques. The aim is to enhance human-machine interaction by enabling intuitive voice control of drone actions. The pipelines developed include: (1) a traditional Speech-to-Text (STT) followed by a Large Language Model (LLM) approach, (2) a direct voice-to-function mapping model, and (3) a Siamese neural network-based system. Each pipeline was evaluated based on inference time, accuracy, efficiency, and flexibility. Detailed methodologies, dataset preparation, and evaluation metrics are provided, offering a comprehensive analysis of each pipeline's strengths and applicability across different scenarios.
\end{abstract}


\section{Introduction}
The integration of automation and voice control in drone systems has received significant attention in recent research, driven by the need for more intuitive and efficient human-machine interaction \cite{hermann2022user, contreras2020unmanned}. This project focuses on developing a voice command system for the Tello drone, utilizing speech recognition and deep learning models to translate voice commands into precise drone actions.

The primary challenge addressed by this project is the accurate and efficient translation of voice commands into specific drone operations. This is particularly crucial in scenarios where traditional control interfaces are impractical or where operators require hands-free operation \cite{yapiciouglu2021voice, krishna2021using}. To address this challenge, we developed and evaluated three distinct pipelines. The first pipeline uses a traditional Speech-to-Text (STT) model followed by a Large Language Model (LLM) for command interpretation \cite{zhao2022improving}. The second pipeline involves a direct mapping model that predicts drone commands from audio inputs without intermediate text conversion. The third pipeline employs a Siamese neural network to generalize new commands by comparing audio inputs to pre-trained examples \cite{xie2021siamese}.

Each pipeline was designed to balance performance, flexibility, and ease of maintenance. The methodologies employed include speech recognition techniques to convert audio to text \cite{krishna2021using}, natural language processing (NLP) for command analysis \cite{de2023semantic}, and neural network models for direct audio-to-command mapping and similarity-based command recognition \cite{xie2021siamese}. The pipelines' effectiveness was evaluated based on accuracy, inference time, and the system's ability to generalize to new commands. The dataset for this project was prepared by recording a variety of voice commands, ensuring diversity in speech patterns and environmental conditions. Data augmentation techniques were applied to enhance model robustness. Evaluation metrics included precision, recall, F1-score, and inference time, providing a comprehensive assessment of each pipeline's performance.

This paper presents a comparative analysis of three voice command pipelines for drone control, highlighting their strengths and potential applications in various operational contexts \cite{hermann2022user, contreras2020unmanned, yapiciouglu2021voice}.

\section{Dataset}
The purpose of the dataset is to provide comprehensive samples of voice commands for controlling the drone. Each command, such as moving the drone "right," "left," "forward," "backward," "up," and "down," was recorded multiple times to capture variations in pronunciation and intonation. 

\subsection{Data Augmentation}
The augmentation step employed various techniques to enhance model robustness and artificially expand the dataset, making it five times larger. This expansion allowed the models to learn more effectively and generalize across diverse scenarios, thereby improving the overall performance and reliability of the voice command system. The data augmentation techniques used were:

\begin{itemize}
    \item \textbf{Noise Addition}: Simulating different environmental sounds to mimic real-world conditions.
    \item \textbf{Tanh Distortion}: Utilizing Tanh activation to normalize audio signals.
    \item \textbf{Masking}: Applying temporal and frequency masking to obscure parts of the audio, promoting generalization.
    \item \textbf{Pitch-Shifting}: Altering the pitch to represent various vocal tones.
\end{itemize}

These techniques ensured that the models could generalize well across different audio conditions. After applying data augmentation, the dataset sizes for each class were increased significantly. The relevant numbers of samples for each class before and after data augmentation are shown in Table \ref{tab:dataset_sizes}.

\begin{table}[h]
    \centering
    \begin{tabular}{|l|c|c|}
        \hline
        \textbf{Class} & \textbf{Without Data Augmentation} & \textbf{With Data Augmentation} \\
        \hline
        RIGHT & 211 & 1055 \\
        BACKWARD & 205 & 1025 \\
        FORWARD & 202 & 1010 \\
        LEFT & 202 & 1010 \\
        UP & 193 & 965 \\
        DOWN & 177 & 885 \\
        \hline
    \end{tabular}
    \vspace{0.15cm}
    \caption{Number of samples per class before and after data augmentation.}
    \label{tab:dataset_sizes}
\end{table}

\section{Methodology}
The development of the voice control system for the Tello drone involved the design and evaluation of three distinct pipelines, each with a specific approach to mapping voice commands to drone actions. The selection and development process of these pipelines were driven by the need to balance performance, flexibility, and ease of maintenance. All three pipelines follow a general workflow to control the Tello drone using voice commands, which involves capturing audio, preprocessing it, and mapping it to specific drone commands, which are then executed. The steps are as follows:

\begin{enumerate}
    \item \textbf{Voice Recording}: Audio is recorded using integrated or external microphones. This step ensures high-quality input suitable for further processing \cite{fayjie2017voice}.
    \item \textbf{Processing}: The captured audio is pre-processed to remove noise and improve signal clarity. This step includes padding the waveform to a consistent length, which is crucial for maintaining uniformity across inputs \cite{meszaros2017speech}.
    \item \textbf{Model Application}: Each pipeline utilizes a different approach to interpret the pre-processed audio and map it to drone commands. The specifics of this step vary across the pipelines:
        \begin{itemize}
            \item Pipeline 1 uses a Speech-to-Text (STT) model followed by a Large Language Model (LLM) \cite{xu2024core}.
            \item Pipeline 2 employs a direct sequence classification model to predict commands from audio inputs \cite{krishna2021using}.
            \item Pipeline 3 utilizes a Siamese neural network to compare audio commands with pre-trained examples \cite{xie2021siamese}.
        \end{itemize}
    \item \textbf{Function Call}: The identified or predicted command is then mapped to a specific function that is executed by the drone \cite{kumaar2018deep}.
\end{enumerate}

We began with Pipeline 1, as it represents the "standard" approach for this type of problem, utilizing Automatic Speech Recognition (ASR) followed by a Large Language Model (LLM) \cite{zhao2022improving}. This approach allowed us to transform voice commands into text and then map that text to specific drone commands. However, we identified opportunities to improve performance and reduce latency by streamlining the process.

This realization led us to develop Pipeline 2, which employs a direct voice-to-function approach. Although this approach is more efficient in terms of latency, it proved less flexible. While ASR and an LLM enable easy addition of new functions, our direct classification model faced challenges in accommodating new commands without significant modifications \cite{krishna2021using}.

To address these limitations, we devised Pipeline 3, which utilizes Siamese neural networks \cite{xie2021siamese}. Siamese networks facilitate the introduction of new functions at a low cost, without necessitating a complete retraining of the model. They project voice commands into a latent space, where similar commands are grouped closely together. This method emerged as the most flexible and efficient solution, offering an ideal balance between performance, flexibility, and ease of maintenance.

\subsection{Preprocessing}
Preprocessing involves several key steps to ensure uniformity and compatibility of the audio data with the models used in all three pipelines. These steps include loading the audio files, standardizing the length of the waveforms, and extracting features to create input values compatible with the Wav2Vec2 model. In all three pipelines, the audio processing involves the following steps:
\begin{itemize}
    \item \textbf{Padding}: Ensuring all waveforms have the same length by adding zeros to the end of waveforms that are shorter than the target length \cite{krishna2021using}.
    \item \textbf{Feature Extraction}: Loading the audio files, standardizing the length of the waveform, and processing the waveform to obtain input values compatible with the Wav2Vec2 model \cite{fayjie2017voice}.
    \item \textbf{Batch Padding}: Padding the input values and labels to ensure that all sequences in a batch have the same length \cite{zhao2022improving}.
\end{itemize}

These steps ensure that the audio data is uniformly processed and compatible with the Wav2Vec2 model used in all pipelines. Additionally, in Pipeline 3, an extra method is implemented to select pairs of audio samples based on the similarity of their classes for audio comparison tasks \cite{xie2021siamese}.

\subsection{Pipeline 1: STT and LLM}
\begin{figure}[htbp]
    \centering
    \includegraphics[width=0.75\textwidth]{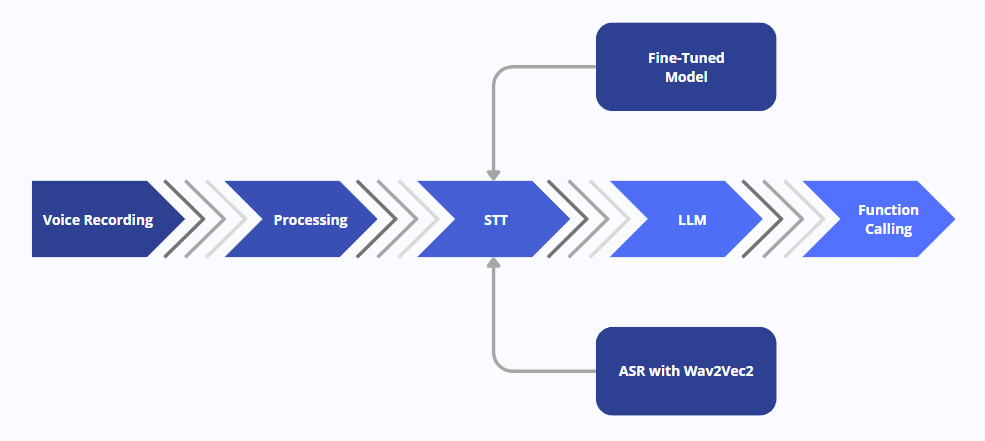}
    \caption{Overview of the STT and LLM pipeline}
    \label{fig:pipeline1}
\end{figure}

The first approach to controlling the Tello drone through voice commands involves converting spoken language into written text using Speech-to-Text (STT) technology, followed by interpreting the text with a Large Language Model (LLM) to generate the appropriate drone commands \cite{xu2024core}.

Speech recognition, or STT, serves as the cornerstone of this pipeline. Our system leverages the "facebook/wav2vec2-large-xlsr-53-portuguese" model developed by Facebook. This model is renowned for its effectiveness in handling sequential data and its robustness against noise and variations in speech. Initially, the model undergoes pre-training on a vast corpus of unlabeled audio data, allowing it to learn general audio features \cite{krishna2021using}. Subsequently, it is fine-tuned with a smaller, labeled dataset to accurately recognize specific speech patterns in Portuguese.

In this pipeline, the audio input is processed and transcribed into text using the wav2vec2 model. Once transcribed, the next step is to interpret the text to discern specific drone commands. We utilize a large language model (LLM) pretrained from Llama3 for this task \cite{xu2024core}. The LLM processes the transcribed text within the context of predefined drone commands: UP, DOWN, FORWARD, BACKWARD, RIGHT, and LEFT. 

To achieve this, we employ a structured approach where the LLM generates responses in JSON format to specify the drone's direction. The process involves a prompt template that provides context about the possible directions the drone can take, and the transcription of the audio command. The response generated by the LLM is structured as a JSON object that specifies the drone's direction, ensuring clear and unambiguous command interpretation.

\subsection{Pipeline 2: Classification Model}
\begin{figure}[htbp]
    \centering
    \includegraphics[width=0.75\textwidth]{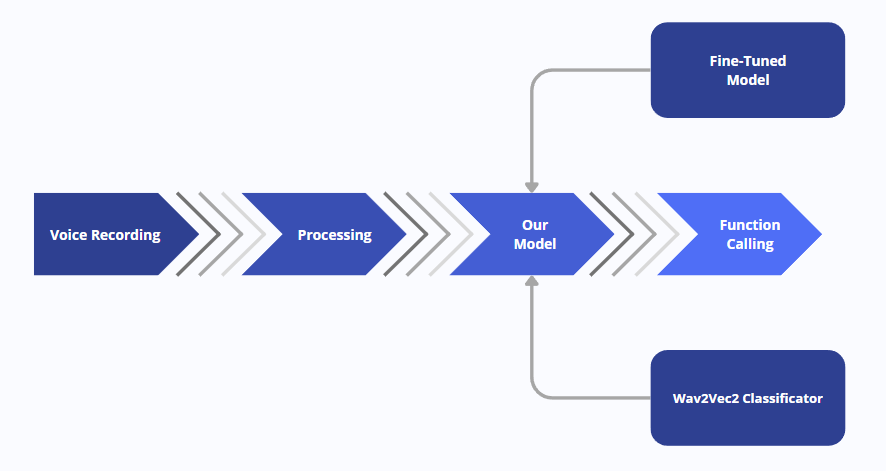}
    \caption{Overview of the Classification Model pipeline}
    \label{fig:pipeline2}
\end{figure}
The second approach involves a direct mapping of audio inputs to drone commands, eliminating the intermediate step of converting speech to text. This pipeline also leverages the "facebook/wav2vec2-large-xlsr-53-portuguese" model, fine-tuned specifically for our project \cite{krishna2021using}. The wav2vec2 model is adept at handling sequential data and effectively processes variations in the audio input.

In this pipeline, audio is captured using integrated or external microphones and undergoes pre-processing to remove noise and improve clarity, including padding the waveform to a consistent length. The pre-processed audio is then fed directly into the wav2vec2 model, which extracts relevant features from the audio waveform and directly predicts the appropriate drone command without requiring a text intermediary \cite{krishna2021using}.

For evaluation, a custom dataset of voice commands was prepared, including audio files and their corresponding labels. The model's performance is assessed on a test set, with key metrics including classification accuracy, mean inference time, and the percentage of unknown commands \cite{krishna2021using}. The evaluation process involves measuring the model's inference time for each prediction and calculating statistical metrics to assess overall performance. A classification report provides detailed insights into the model's accuracy across different command classes.

By using the same base model, the "facebook/wav2vec2-large-xlsr-53-portuguese", both pipelines demonstrate different methodologies for drone command recognition. Pipeline 1 uses an intermediate STT step followed by LLM for command interpretation, while Pipeline 2 employs direct audio-to-command mapping, showcasing the versatility and robustness of the wav2vec2 model in handling various tasks \cite{krishna2021using}.

\subsection{Pipeline 3: Siamese Network}
\begin{figure}[htbp]
    \centering
    \includegraphics[width=0.75\textwidth]{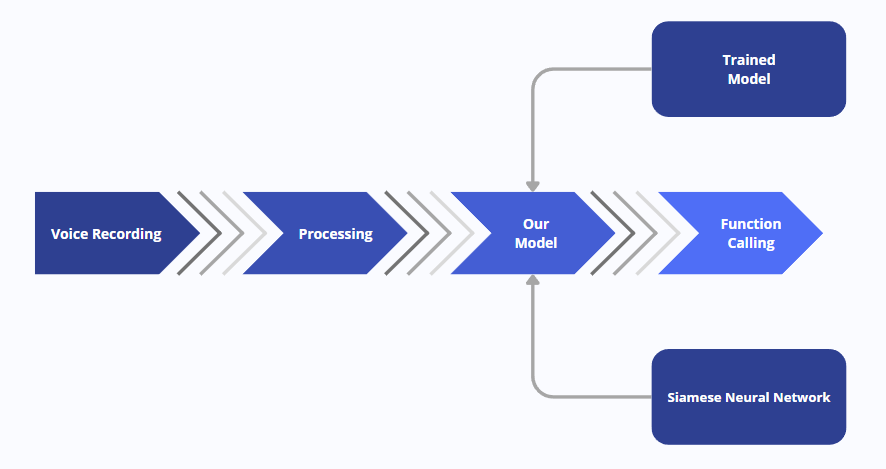}
    \caption{Overview of the Siamese Network pipeline}
    \label{fig:pipeline3}
\end{figure}

The third approach explores the use of Siamese neural networks, which are particularly adept at comparing pairs of inputs to determine their similarity \cite{xie2021siamese}. This characteristic is invaluable for tasks that require the system to generalize to new examples, such as recognizing new voice commands without necessitating extensive retraining.

A Siamese network consists of two or more identical sub-networks that share the same weights and architecture. During training, pairs of examples are fed into these sub-networks. The network learns to project similar examples close to each other in a latent space while ensuring that dissimilar examples are projected farther apart \cite{xie2021siamese}. This process is facilitated by the use of contrastive loss, which minimizes the distance between similar examples and maximizes the distance between dissimilar ones. This training approach enables the network to effectively distinguish between various commands based on their latent representations \cite{xie2021siamese}.

In our system, the Siamese network includes multiple convolutional layers to extract features from the input audio signals. The network is trained to encode audio files into fixed-size vectors. These vectors are stored in a vector database, where each vector is associated with a corresponding command label \cite{fayjie2017voice}. When a new voice command is received, it is encoded into a vector and compared against the stored vectors using a K-Nearest Neighbors (KNN) model. The KNN model identifies the closest matching command based on the vector similarity, thus determining the appropriate action for the drone \cite{fayjie2017voice}.

This methodology allows for flexible and efficient recognition of new commands, significantly reducing the need for extensive retraining and enabling quick adaptation to new instructions. By leveraging the Siamese network's ability to generalize from a relatively small set of training examples, we achieve robust and scalable voice command recognition for the Tello drone.

\section{Results}
In this section, we present the results obtained after training and evaluating the proposed pipelines. The results include accuracy, precision, recall, F1-score, inference times, and a summary comparison of all three pipelines.

\subsection{Pipeline 1: STT and LLM}
The first pipeline involves converting spoken language into written text using Speech-to-Text (STT) technology, followed by interpreting the text with a Large Language Model (LLM) to generate the appropriate drone commands. The performance metrics for this pipeline are presented in Table \ref{tab:results_pipeline1}.

\begin{table}[ht]
    \centering
    \begin{tabular}{|c|c|c|}
        \hline
        Metric & Without Fine-Tuning & With Fine-Tuning \\
        \hline
        Accuracy & 0.73 & 0.81 \\
        Precision & 0.74 & 0.78 \\
        Recall & 0.63 & 0.70 \\
        F1-Score & 0.66 & 0.73 \\
        \hline
    \end{tabular}
    \vspace{0.15cm}
    \caption{Performance Metrics for Pipeline 1}
    \label{tab:results_pipeline1}
\end{table}

The accuracy and precision metrics indicate a significant improvement with fine-tuning. The model achieved an accuracy of 0.81 and an F1-score of 0.73 after fine-tuning, demonstrating its capability to accurately transcribe and interpret voice commands.

\subsection{Pipeline 2: Direct Model}
The second pipeline involves a direct mapping of audio inputs to drone commands, bypassing the intermediate step of converting speech to text. The performance metrics for this pipeline are presented in Table \ref{tab:results_pipeline2}.

\begin{table}[ht]
    \centering
    \begin{tabular}{|c|c|c|}
        \hline
        Metric & Without Fine-Tuning & With Fine-Tuning \\
        \hline
        Accuracy & 0.94 & 0.99 \\
        Precision & 0.91 & 0.98 \\
        Recall & 0.93 & 0.99 \\
        F1-Score & 0.92 & 0.98 \\
        \hline
    \end{tabular}
    \vspace{0.15cm}
    \caption{Performance Metrics for Pipeline 2}
    \label{tab:results_pipeline2}
\end{table}

This pipeline also shows notable improvements with fine-tuning, achieving an accuracy of 0.99 and an F1-score of 0.98. The direct classification approach proves to be effective in interpreting voice commands without the need for an intermediate text representation.

\subsection{Pipeline 3: Siamese Network}
The third pipeline utilizes a Siamese neural network to compare pairs of inputs and determine their similarity. The performance metrics for this pipeline are presented in Table \ref{tab:results_pipeline3}.

\begin{table}[ht]
    \centering
    \begin{tabular}{|c|c|c|}
        \hline
        Metric & Without Fine-Tuning & With Fine-Tuning \\
        \hline
        Accuracy & 0.54 & 0.74 \\
        Precision & 0.54 & 0.74 \\
        Recall & 0.54 & 0.75 \\
        F1-Score & 0.54 & 0.74 \\
        \hline
    \end{tabular}
    \vspace{0.15cm}
    \caption{Performance Metrics for Pipeline 3}
    \label{tab:results_pipeline3}
\end{table}

The Siamese network approach showed significant improvement with fine-tuning, achieving an accuracy of 0.74. The ability to generalize to new commands was enhanced, making this pipeline particularly useful for scenarios requiring flexibility and adaptability.

\subsection{Summary of Results}
Table \ref{tab:summary_results} provides a summary comparison of the accuracy and inference times for all three pipelines.

\begin{table}[ht]
    \centering
    \begin{tabular}{|c|c|c|}
        \hline
        Pipeline & Accuracy & Inference Time (in seconds) \\
        \hline
        STT and LLM & 0.81 & 1.233 \\
        Direct Model & 0.99 & 0.021 \\
        Siamese Network & 0.74 & 0.006 \\
        \hline
    \end{tabular}
    \vspace{0.15cm}
    \caption{Summary of Results for All Pipelines}
    \label{tab:summary_results}
\end{table}

The results indicate that the Direct Model pipeline achieved the highest accuracy of 0.99. However, the Siamese Network pipeline significantly outperformed the others in terms of inference time, making it the most efficient for real-time applications. Despite being slightly lower in accuracy at 0.74, the Siamese Network demonstrated the best performance in inference time and the ability to generalize to new commands.

\subsection{Discussion}
The evaluation of the three pipelines highlights the trade-offs between accuracy, inference time, and flexibility. The STT and LLM pipeline, while highly accurate, has a longer inference time due to the sequential nature of the tasks involved. The Direct Model pipeline provides the highest accuracy and a balance between precision and efficiency, making it highly suitable for real-time applications. The Siamese Network pipeline offers the best generalization capabilities and the shortest inference time, which is advantageous in dynamic environments where new commands might be introduced.

Future work will focus on further improving the models, expanding the dataset, and exploring additional techniques to enhance the performance and reliability of the voice command system for drone control.

\section{Conclusions}
The development and evaluation of the three voice command pipelines for controlling the Tello drone demonstrate the effectiveness of using different approaches: STT followed by LLM, direct classification, and Siamese networks. Each pipeline has its unique strengths and potential applications, depending on the specific requirements of inference time, accuracy, efficiency, and flexibility. Through a comparative analysis, we have highlighted the trade-offs between these approaches.

The results indicate that Pipeline 1 (STT and LLM) showed high accuracy and precision, but with a longer inference time compared to other pipelines. Pipeline 2 (Direct Model) proved to have the highest accuracy and precision, with a balance between precision and efficiency. Pipeline 3 (Siamese Network) showed promise in generalizing to new commands, offering the best inference time and flexibility. These findings suggest that the ideal pipeline choice depends on the specific application context and requirements. In situations requiring high precision, Pipeline 2 is most suitable, whereas Pipeline 3 is preferable where speed and flexibility are crucial. Pipeline 1 offers a balanced solution, especially useful for applications requiring high accuracy in command interpretation.

The implementation of this voice-controlled drone system demonstrates the potential of utilizing STT, NLP, and LLM technologies to create intuitive and efficient interfaces for drones. In the future, improving models and collecting more extensive datasets can further enhance the system's performance and applicability.


\begin{thebibliography}{10}

\bibitem{contreras2020unmanned}
Ruben Contreras, Angel Ayala, and Francisco Cruz.
\newblock Unmanned aerial vehicle control through domain-based automatic speech
  recognition.
\newblock {\em Computers}, 9(3):75, 2020.

\bibitem{de2023semantic}
J~De~Curt{\'o}, I~De~Zarza, and Carlos~T Calafate.
\newblock Semantic scene understanding with large language models on unmanned
  aerial vehicles.
\newblock {\em Drones}, 7(2):114, 2023.

\bibitem{fayjie2017voice}
Abdur~Razzaq Fayjie, Amir Ramezani, Doukhi Oualid, and Deok~Jin Lee.
\newblock Voice enabled smart drone control.
\newblock In {\em 2017 Ninth international conference on ubiquitous and future
  networks (ICUFN)}, pages 119--121. IEEE, 2017.

\bibitem{hermann2022user}
Julia Hermann, Moritz Pl{\"u}ckthun, Ayseg{\"u}l Dogang{\"u}n, and Marc
  Hesenius.
\newblock User-defined gesture and voice control in human-drone interaction for
  police operations.
\newblock In {\em Nordic Conference on Human-Computer Interaction}, pages
  1--11, 2022.

\bibitem{krishna2021using}
DN~Krishna, Pinyi Wang, and Bruno Bozza.
\newblock Using large self-supervised models for low-resource speech
  recognition.
\newblock In {\em Interspeech}, pages 2436--2440, 2021.

\bibitem{kumaar2018deep}
Saumya Kumaar, Toshit Bazaz, Sumeet Kour, Disha Gupta, Ravi~M Vishwanath,
  SN~Omkar, et~al.
\newblock A deep learning approach to speech based control of unmanned aerial
  vehicles (uavs).
\newblock In {\em CS \& IT Conf. Proc}, volume~8, 2018.

\bibitem{meszaros2017speech}
Erica~L Meszaros, Meghan Chandarana, Anna Trujillo, and B~Danette Allen.
\newblock Speech-based natural language interface for uav trajectory
  generation.
\newblock In {\em 2017 International Conference on Unmanned Aircraft Systems
  (ICUAS)}, pages 46--55. IEEE, 2017.

\bibitem{xie2021siamese}
Yang Xie, Zhenchuan Zhang, and Yingchun Yang.
\newblock Siamese network with wav2vec feature for fake speech detection.
\newblock In {\em Interspeech}, pages 4269--4273, 2021.

\bibitem{xu2024core}
Shuyuan Xu, Zelong Li, Kai Mei, and Yongfeng Zhang.
\newblock Core: Llm as interpreter for natural language programming,
  pseudo-code programming, and flow programming of ai agents.
\newblock {\em arXiv preprint arXiv:2405.06907}, 2024.

\bibitem{yapiciouglu2021voice}
Cengizhan Yapicio{\u{g}}lu, Z{\"u}mray Dokur, and Tamer {\"O}lmez.
\newblock Voice command recognition for drone control by deep neural networks
  on embedded system.
\newblock In {\em 2021 8th International Conference on Electrical and
  Electronics Engineering (ICEEE)}, pages 65--72. IEEE, 2021.

\bibitem{zhao2022improving}
Jing Zhao and Wei-Qiang Zhang.
\newblock Improving automatic speech recognition performance for low-resource
  languages with self-supervised models.
\newblock {\em IEEE Journal of Selected Topics in Signal Processing},
  16(6):1227--1241, 2022.

\end{thebibliography}

\end{document}